\documentclass[draft,jgrga]{agutex}

 \usepackage{lineno}
 \usepackage{multirow}
 \linenumbers*[1]

  \usepackage[dvips]{graphicx}
  \usepackage{color}
\usepackage{amssymb,amsmath} 
  \setkeys{Gin}{draft=false}
\authorrunninghead{QIN AND ZHAO}

\titlerunninghead{Solar Cycle variations of SEPs and CRs}

\authoraddr{G. Qin,
State Key Laboratory of Space Weather, Center for Space Science and
Applied Research, Chinese Academy of Sciences, P.O. Box 8701,
Beijing 100190, China, (gqin@spaceweather.ac.cn)}

\authoraddr{L.-L. Zhao,
State Key Laboratory of Space Weather, Center for Space Science and
Applied Research, Chinese Academy of Sciences, P.O. Box 8701,
Beijing 100190, China, (llzhao@spaceweather.ac.cn)}

\begin{document}

%
%

\title{Study of Different Solar Cycle Variations of Solar Energetic Particles and 
Cosmic Rays by Despiking ACE/SIS Heavy-Ion Fluxes}
%

%
%



\authors{G. Qin, and L.-L. Zhao}
\affil{State Key Laboratory of Space Weather,
Center for Space Science and Applied Research, Chinese Academy of Sciences,
P.O. Box 8701, Beijing 100190, China}





%
%


\begin{abstract}
Cosmic Rays (CRs) include Galactic Cosmic Rays (GCRs) and Anomalous Cosmic Rays 
(ACRs). The CR flux data of protons and heavy-ions observed with spacecraft are 
often seriously contaminated by Solar Energetic Particle (SEP) events. In this work,
 we separate SEPs from CRs of ACE/SIS spacecraft observations with an automatic 
despiking algorithm, so we are able to study the 
different variations of SEPs and CRs over a solar cycle. In particular, we study the 
 elemental ratio, first ionization potential dependence and elemental dependence, 
and information entropy of SEPs and CRs. So that we can gain new insights into
 energetic particles' different compositions, origins, and transport processes, etc.
\end{abstract}

%
%

%

\begin{article}

%
%

\section{Introduction}
The Solar Energetic Particles (SEPs) are high energy particles produced by solar 
events with abrupt variations with the process of solar events. In addition,
the Cosmic Rays (CRs), including Galactic Cosmic Rays (GCRs) and
Anomalous Cosmic Rays (ACRs), of protons and heavy-ions can be assumed as a 
background gradually evaluating over the solar cycle. Therefore, CRs and SEPs have 
distinctively different behaviours varying with solar activities.
Furthermore, ACRs and GCRs are also obviously different, since ACRs are usually 
believed to be accelerated by termination shock of heliosphere, while GCRs have 
sources outside the heliosphere.

However, GCRs from spacecraft observations are often significantly 
contaminated by SEP events, especially during solar maximum. 
Recently, \citet{Qin2012APJ} used a robust 
despiking algorithm based on the Poincare map 
thresholding method \citep{goring2002} to eliminate the spiked flux records possibly
 associated with SEP events from continuous GCR records. Based on this method, 
\citet{Zhao2013JGR} and \citet{ZhaoEA2013} used the despiking algorithm to remove 
the spikes of ACE/SIS 
flux to obtain the clean background of CRs. Furthermore, using the GCR data from 
ACE/CRIS, \citet{Zhao2013JGR} developed an observation-based elemental GCR heavy-ion
 spectra model. It is shown that the model, if extrapolated, agrees with the GCR 
data from ACE/SIS with despiking very well.

\citet{Reames2013} studied the abundances of the heavy elements in SEPs with energy
$2-15$ MeV/nuc measured on Wind spacecraft during $54$ large SEP events for the
peorid over $17$ years. It is shown that the coronal abundances 
can be determined from SEP measurements, and their value relative to those in the 
solar photosphere depend on the first ionization potential (FIP) of the element.
Furthermore,
Shannon entropy (also named as information entropy) is a commonly used quantity 
to characterize the degree of uncertainty and/or the information contained in a 
time-series signal. \cite{Laurenza2012} used information entropy as a proxy of 
changes of spectrum shape in a limited energy range to study the evolution of the 
differential flux spectrum during an SEP event. By investigating the time profile
of Shannon entropy during an SEP event, they concluded that a perpendicular shock in
the solar corona produced the particle acceleration of the event.

In the study of SEPs, usually fluxes in different phases of individual SEP events 
are studied carefully. However, in this work, we demonstrate that it is feasible to 
study the different characteristics of SEPs and CRs (GCRs or ACRs), by separating 
their records in spacecraft observations over solar cycles automatically with the 
aforementioned despiking algorithm. The article is organized as follows. We discuss
the despiking of ACE/SIS heavy ions to separate CRs from SEPs in section 2. The 
elemental fluxes and their ratios for SEPs and CRs are discussed in section 3. The
first ionization potential dependence of SEP fluxes and elemental dependence of CR
fluxes are discussed in subsections 4.1 and 4.2, respectively. The shannon entropy 
of CRs and SEPs from heavy ions are shown in section 5. Finally, discussion and 
conclusion are presented in section 6.

\section{Data and Despiking Algorithm}
The NASA Advanced Composition Explorer (ACE) spacecraft has greatly extended our 
ability to explore the heavy nuclei over a wide energy range \citep{Stone1998SSRa}, 
and the flux measurements from ACE spacecraft are the most reliable and 
statistically significant heavy-ion data so far, due to their large geometrical 
acceptance and excellent charge and mass resolution \citep{George2009APJ}.

In this study, we use heavy-ion flux data from the Solar Isotope Spectrometer (SIS) 
instrument onboard ACE spacecraft. The SIS instrument is designed to provide high 
resolution measurements of the isotopic composition of energetic nuclei from He to 
Ni over the energy range $10$ MeV/nuc - $100$ MeV/nuc. It records SEPs during large 
solar events, and CRs (GCRs or ACRs) during solar quiet time, so we are provided a 
unique opportunity to compare the different characteristics between SEPs and CRs 
(GCRs or ACRs).

\begin{figure}
  \centering
  \includegraphics[width=0.7\textwidth]{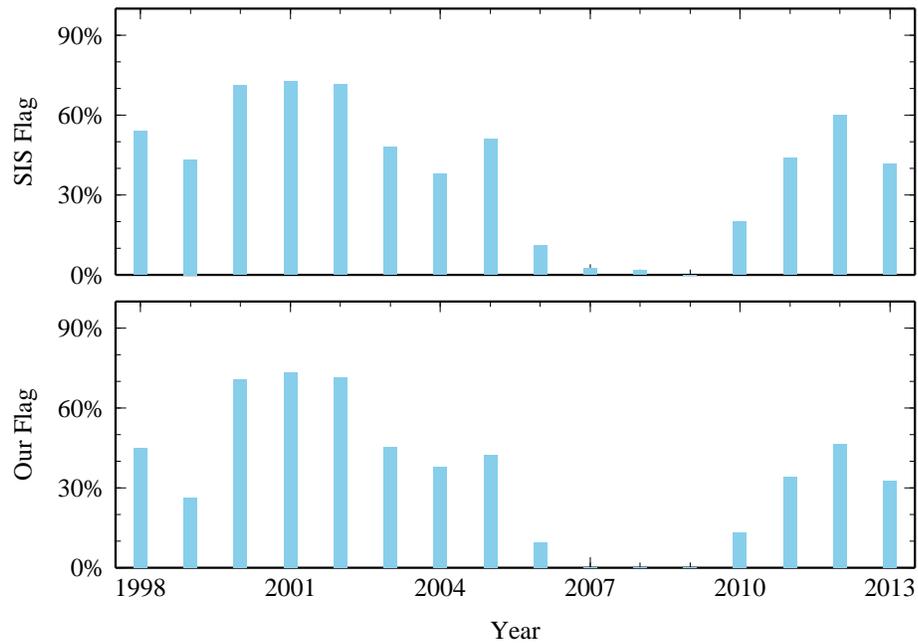}
  \caption{The percentage of flagged periods in each year with SIS flag (top panel)
and our flag (bottom panel).}
  \label{fig:flag}
\end{figure}

The flux measurements for heavy elemental species (atomic number $2\le z\le 28$) of 
the SIS (if available) at level 2 over the period from the year 1997 to 2013 are 
readily obtained from the ACE Science Center, with recommended energy ($E_i$) in 
unit of MeV/nuc of each energy interval shown in Table \ref{tbl:energy_interval}. 
Then, we use He(2) data of SIS 
measurements to flag periods of spikes, possibly associated with SEP events, with a
 despiking algorithm \citep{Qin2012APJ} \citep[see also,][]{Zhao2013JGR} based on 
the Poincare Map Thresholding method \citep{goring2002}. Specifically, we go through 
the last five SIS He channels ( $>8$ MeV/nuc), at any time if one of the He channels 
is considered as a spike with the despiking algorithm, the period is flagged as 
solar activity. Here, the Universal threshold $\lambda_U=1.1$ in the thresholding
algorithm \citep{Qin2012APJ}. Note 
that the data from SIS instrument include a ``solar activity flag'' 
too (noted as ``SIS flag'' hereafter) that flags periods with significant SEP 
contributions. Figure \ref{fig:flag} illustrates the percentage of flagged 
one-day-periods in each year from year 1998 to year 2013 with SIS flag (top panel) 
and our flag (bottom panel). The two panels of Figure \ref{fig:flag} generally show 
similar variation patterns. In addition, we investigate the flagged 
one-hour-periods and get an ``event'' for each consecutive range of flagged 
periods. Here, we assume an event as an impulsive one if it lasts less than a day, 
otherwise, it is assumed as a gradual one. We find that of all the events only a few
 are impulsive ones, so this method is only good at distinguish 
gradual solar events from CRs background. In the next, we get our flag using the 
despiking algorithm \citep{Qin2012APJ} with a computer automatically and discard the
 impulsive events and divide the rest data into two groups, the gradual SEP data 
during the flagged periods and the CR data during the non-flagged periods. 

\begin{figure}
  \centering
  \includegraphics[width=0.7\textwidth]{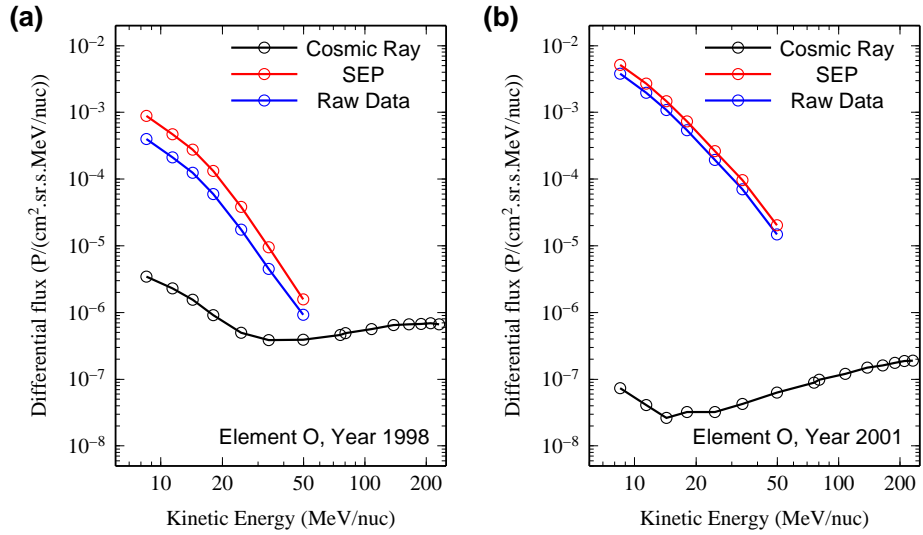}
  \caption{The kinetic energy spectra of element O for year 1998 (left panel) and 
year 2001 (right panel). The blue, red, and black lines indicate raw data from SIS 
measurements, SEP data from flagged periods, and CR background data from the
non-flagged periods, respectively.}
  \label{fig:spectra}
\end{figure}

Figure \ref{fig:spectra} shows the kinetic energy spectra of element O for the year
1998 (left panel, solar minimum) and the year 2001 (right panel, solar maximum). The
 blue lines indicate raw data from SIS measurements, the red lines indicate SEP data
 from flagged periods, and the black lines indicate CR background data from the 
non-flagged periods. From the figure we can see, that during the solar minimum the
CR data are much higher than that during the solar maximum, but the SEP data are 
much lower than that during the solar maximum. In addition, for the CR data, the
energy range with positive spectral index indicates GCR, while that with negative
spectral index indicates ACR, so it is shown that the upper limit of ACR dominate 
energy range is much larger during the solar minimum.

\section{Flux Ratio for SEP and CR Data}
The SIS instrument measures the differential flux of each element over $8$ different
 consecutive energy intervals. In order to study flux ratio for SEP and CR data, we 
choose three energy channels, I, II, and III with energy approximately equal to 
$13$, $31$, and $46$ MeV/nuc, respectively, for each of N, O, and Fe as shown in 
Table \ref{tbl:channels}. In Table \ref{tbl:channels}, $E_i$ shows the energy 
channel for each of
N, O, and Fe shown in Table \ref{tbl:energy_interval}. Note that the measurements of
 $13$ MeV/nuc channels of N and O are considered mainly from ACRs, that the flux of
 $31$ MeV/nuc channels of N and O are considered dominated with GCRs with some 
contribution from ACRs, and that of the rest channels are mainly from GCRs with
much less ACRs contribution.

\begin{figure}
  \centering
  \includegraphics[width=.4\textwidth]{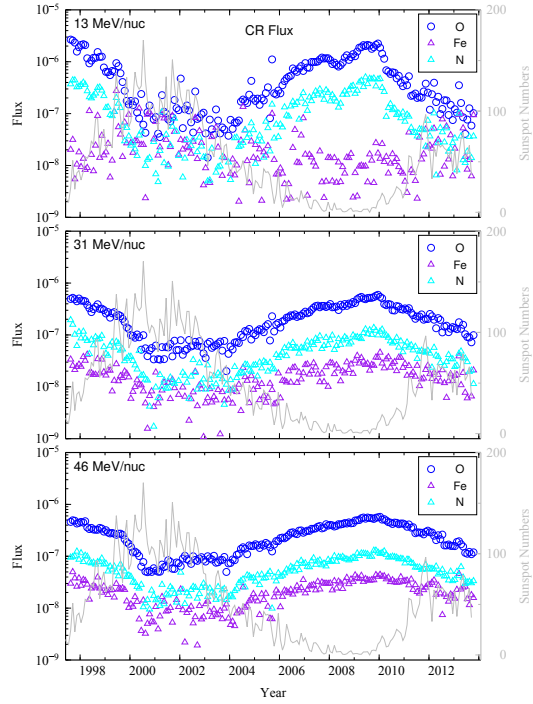}
  \caption{
   CR fluxes of O (blue circles), Fe (purple triangles), and N (cyan triangles), 
corresponding to the energy channels $13$ MeV/nuc (top pannel), $31$ MeV/n (middle 
pannel), and $46$ MeV/nuc (bottom panel). Note that the grey line represents the 
time variations of sunspot number. 
  }
  \label{fig:ratio-2}
\end{figure}

Figure \ref{fig:ratio-2} shows despiked monthly fluxes, or CRs, of O (blue
circles), Fe (purple triangles), and N (cyan triangles) with energy channels $13$ 
MeV/nuc (top pannel), $31$ MeV/nuc (middle pannel), and $46$ MeV/nuc (bottom panel). 
Note that the grey line indicates the time variations of sunspot numbers (SSN). It 
is shown that for all the O, N, and Fe channels the CR fluxes are anti-correlated 
with SSN. In addition, O and N channels show stronger anti-correlation than Fe 
channels, especially for the lowest energy channel, $13$ MeV/nuc.

\begin{figure}
  \centering
  \includegraphics[width=.8\textwidth]{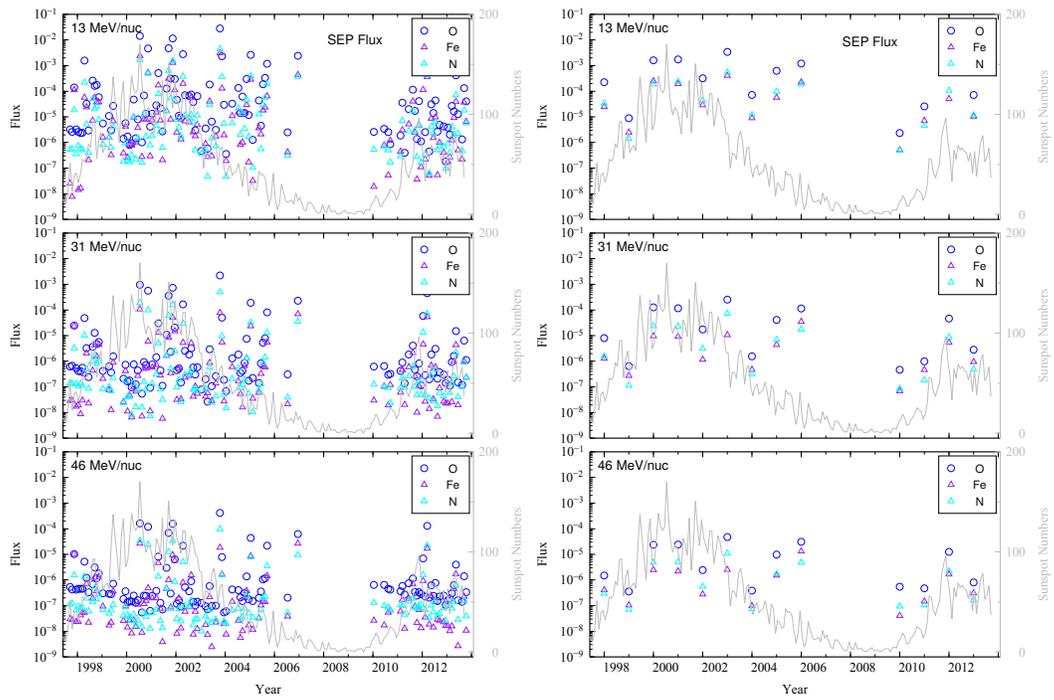}
  \caption{
    Similar as Figure \ref{fig:ratio-2} except that color symbols indicate SEP 
fluxes in monthly and yearly average in left and right panels, respectively.
  }
  \label{fig:ratio-3}
\end{figure}

Figure \ref{fig:ratio-3} is similar as Figure \ref{fig:ratio-2} except that the 
color symbols indicate SEP fluxes in monthly and yearly average in left and right
panels, respectively. It is shown that the SEP fluxes have much larger
variations in any epoch of solar activity, which is a typical behavior for gradual
SEPs \citep{Reames2013}. In addition, the SEP fluxes also show moderate correlation 
with solar cycles, and the correlation is stronger for lower energy channels. In 
addition, the fluxes are larger during solar maximum and descending phases than that
 during solar minimum and ascending phases.

\begin{figure}
  \centering
  \includegraphics[width=.8\textwidth]{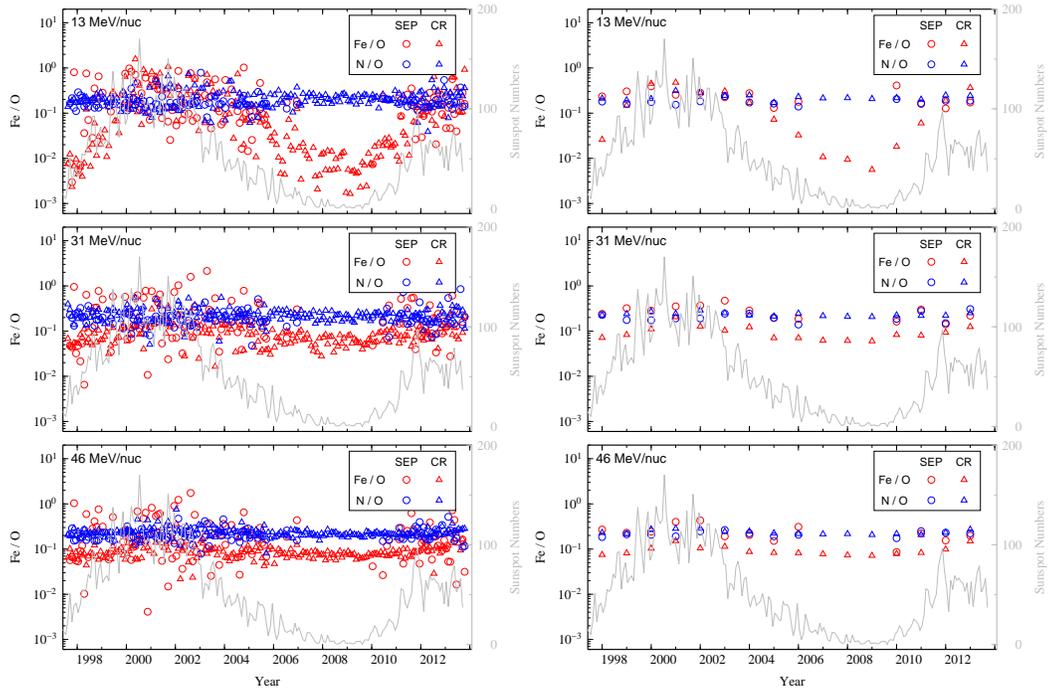}
  \caption{
    Similar as Figure \ref{fig:ratio-3} except that red and blue symbols indicate 
Fe/O ratio and N/O ratio, respectively, and that circles and triangles indicate SEPs
 and CRs, respectively.
  }
  \label{fig:ratio-1}
\end{figure}

Figure \ref{fig:ratio-1} is similar as Figure \ref{fig:ratio-3} except
that red and blue symbols indicate Fe/O ratio and N/O ratio, respectively, and that
circles and triangles indicate SEPs and CRs, respectively. From the figure we can 
see the Fe/O ratio of CRs in monthly average is obviously correlated
to the solar activity in lower two energy channels, but the correlation in the 
highest energy channel is much smaller, indicating different modulation process of 
ACRs and GCRs under varied modulation strength. In addition, it is found 
that the Fe/O ratio of SEPs have very large variations in any solar cycle 
epoch. Although the distribution of Fe/O ratio of SEPs are scattering, the ratio is 
distinctly higher than that of CR.
However, the N/O ratio of SEPs shows much less variations in any 
solar cycle epoch, and both of CRs and SEPs of N/O don't show much solar activity 
correlation. In addition, the yearly average of N/O ratio of CRs is usually larger 
than that of SEPs. Furthermore, the ratio N/O of CRs and SEPs are generally larger 
than the ratio Fe/O of CRs and SEPs, respectively.

For the ratio of Fe/O of $13$ MeV/nuc CR particles, the apparent solar cycle 
variations are caused by the strong solar cycle dependence of ACR flux of O and weak
solar cycle dependence of GCR flux of Fe. And for the ratio of Fe/O of $31$ MeV/nuc
CR particles, there is similar but much weaker solar cycle dependence because in 
$31$ MeV/nuc O channel, ACRs are not dominated as that in $13$ MeV/nuc O channel,
and the solar cycle dependence of $31$ MeV/nuc O CRs is relatively weaker and 
the solar cycle dependence of $31$ MeV/nuc Fe CRs is relatively stronger 
than that with energy $13$ MeV/nuc. 

\section{First Ionization Potential Dependence of SEP Fluxes and Elemental 
Dependence of GCR Fluxes}

In order to obtain the flux at any energy per nucleon, $E$, for any heavy nuclei 
$n$, we get the energy channels $E_i$ and $E_{i+1}$ with $E_i\le E\le E_{i+1}$ in 
Table \ref{tbl:energy_interval}. For each year $t$ from 1998 to 2013 we fit the 
yearly averaged SEP and CR fluxes of each heavy ACE/SIS nuclei $n$ from
channel $E_i$ to $E_{i+1}$ shown in Table \ref{tbl:energy_interval} with
$f_{SEP}^F(n,E,t)=f_{SEP~0}(n,t) E^{k(n,t)}$ and 
$f_{CR}^F(n,E,t)=f_{CR~0}(n,t) E^{q(n,t)}$, respectively. From the 
fitting result, we can get the SEP at $E=15$ MeV/nuc and GCR at $E=30$ 
MeV/nuc, for each year and
 element, $f_{SEP}^F(n,15\text{~MeV/nuc},t)$ and $f_{CR}^F(n,30\text{~MeV/nuc},t)$,
respectively. Note that from Figure \ref{fig:flag} we can see in the years 1999 and
2006 to 2010, the percentage of flagged periods with our flag is all smaller than
$30\%$, so below we do not get fitting of SEPs in these years because of too few SEP 
events. In next two subsections, we study different dependence of SEP fluxes
of $15$ MeV/nuc and GCR fluxes of $30$ MeV/nuc.

\subsection{First Ionization Potential Dependance of SEP Fluxes Relative to 
Photospheric Abundances}

\citet{Reames2013} showed that SEP abundances divided by recent photospheric 
abundances can be approximately represented as a power law of First Ionization 
potential (FIP).

\begin{figure}
  \centering
  \includegraphics[width=0.7\textwidth]{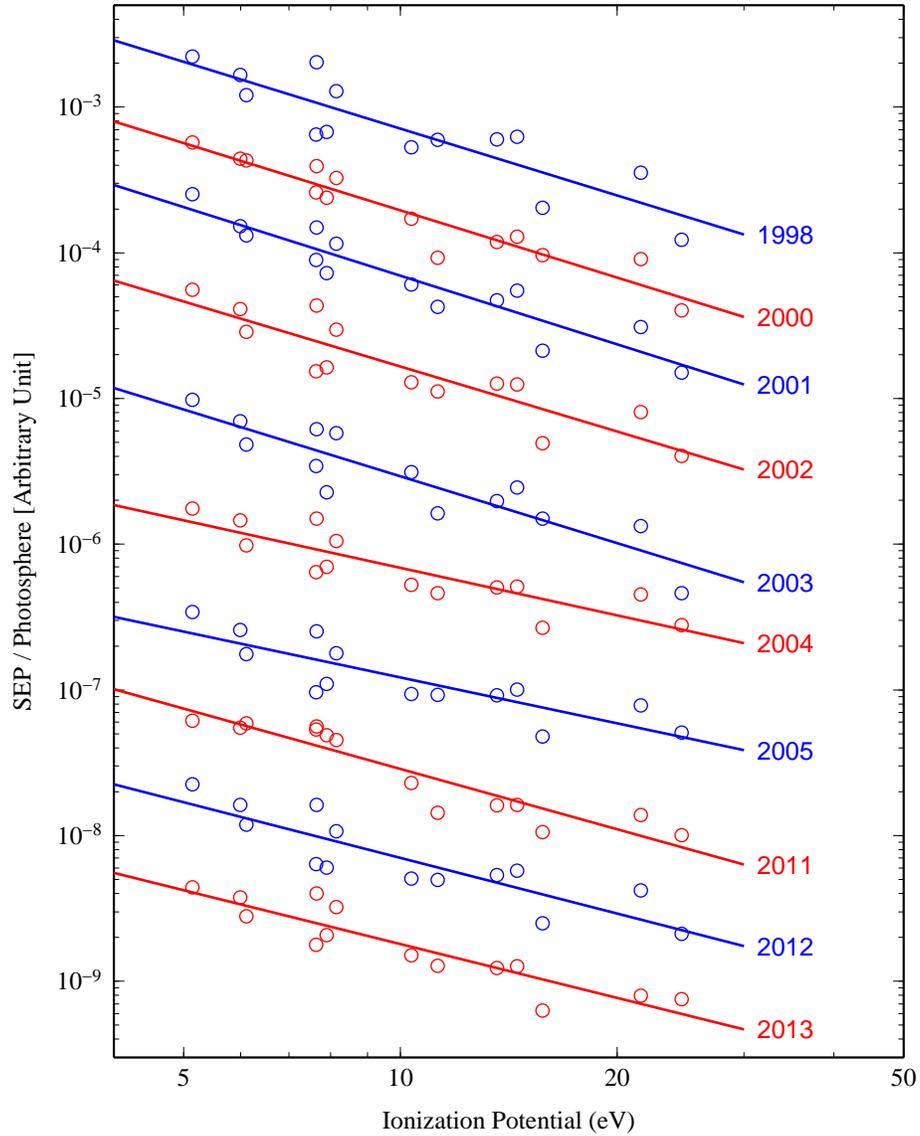}
  \caption{Fitted $15$ MeV/nuc element fluxes for yearly average of SEP events 
relative to the recent photospheric abundances are shown as a function of FIP. The 
results are multiplied by a free number for presentation purpose. The solid lines are
 least-square linear fit in log-log space.}
  \label{fig:fip}
\end{figure}

Figure \ref{fig:fip} illustrates the fitted yearly SEP fluxes at $15$ MeV/nuc
divided by recent photospheric abundances as a function of FIP from the year 1998 
to 2013, except the years 1999, and 2006-2010, in arbitrary unit. Note that the 
results are multiplied 
by a free parameter for the purpose of presentation and that the recent photospheric 
abundances of elements and FIP are taken from Table 1 of \citet{Reames2013}. It is 
shown that the $15$ MeV/nuc SEP fluxes divided by photospheric abundances are 
linearly correlated with FIP in log-log space, with fitted slopes, intercepts, and
correlation coefficients. 

\begin{figure}
  \centering
  \includegraphics[width=0.7\textwidth]{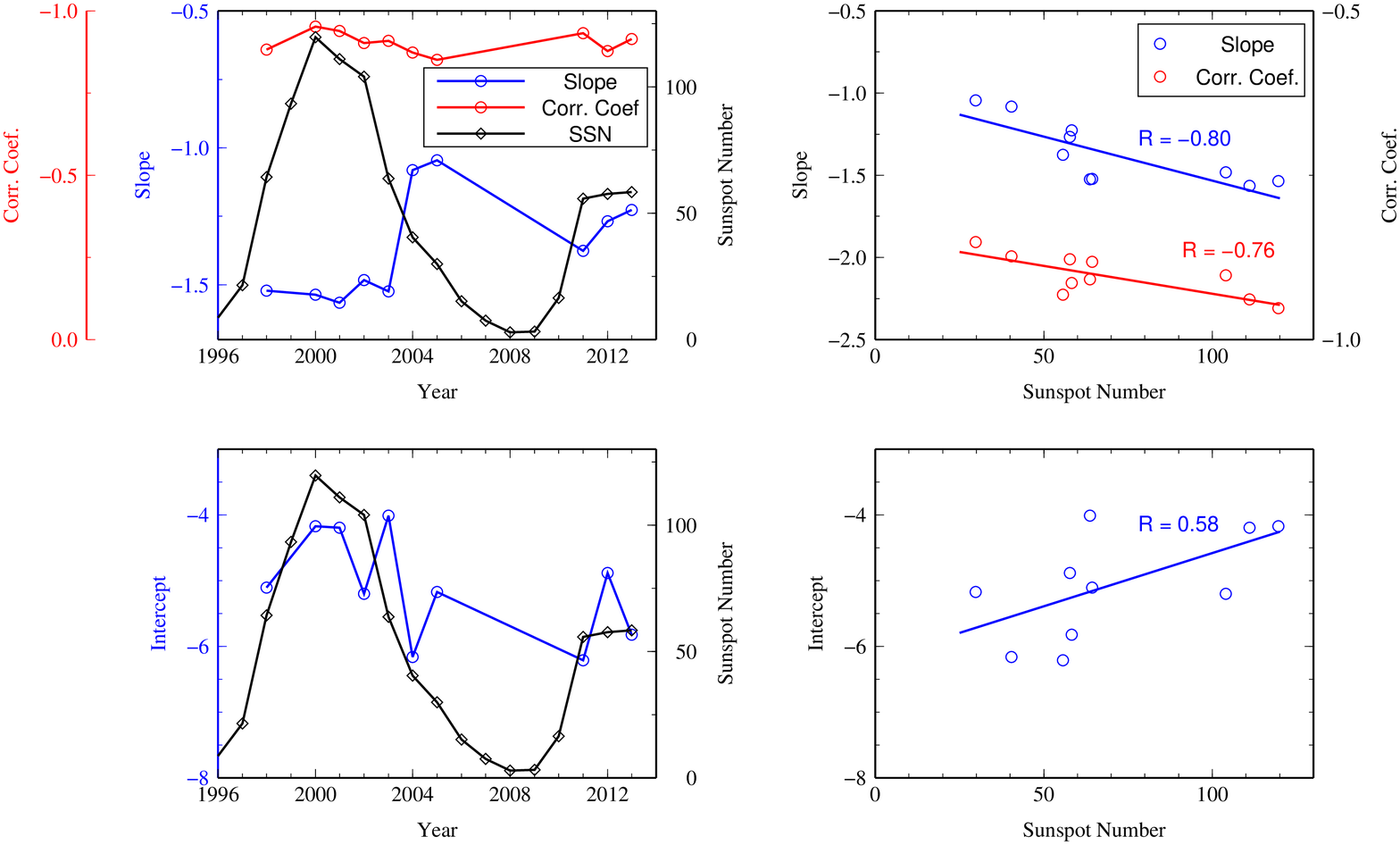}
  \caption{The slope, intercept, and correlation coefficients (Corr. Coef.) of the 
least-square linear fit in Figure \ref{fig:fip} are shown. Top left panel shows the 
slope, Corr. Coef., and sunspot numbers (SSN) varying as the year. Top right panel  
shows the slope and Corr. Coef. varying as the SSN, and linear-fit of the slope and
Corr. Coef. as functions of SSN are also shown. Bottom left panel shows the 
intercept and SSN varying as the year. Bottom right panel shows the intercept as a 
function of the SSN, and the linear-fit is also shown.}
  \label{fig:slope}
\end{figure}

In Figure \ref{fig:slope} we show the fitted slopes, correlation coefficients (Corr. 
Coef.), and intercepts of the above linear relationship. Top left panel of the 
figure shows the slopes, Corr. Coef., and sunspot numbers (SSN) varying as the year.
 It is shown that the Corr. Coef. of the linear relationship in Figure \ref{fig:fip}
 are all near $-1$, indicating a strong anti-correlation. And
the top right panel shows the slopes and Corr. Coef. varying as the SSN, we can see
when SSN is large, the Corr. Coef. is near $-1$, indicating stronger 
linear-relationship in Figure \ref{fig:fip}. Furthermore, the slopes and Corr. Coef. 
can be linearly fitted as function of SSN with 
fitting correlation coefficients $R=-0.80$ and $R=-0.76$, respectively. Moreover,
the bottom left panel shows the intercepts and SSN varying as the year. In addition, 
the bottom right panel shows the intercepts as a function of the SSN, and the 
intercepts can also be linearly fitted as a function of SSN with fitting correlation 
coefficient $R=0.58$.

\subsection{GCR Abundance Relative to Atomic Numbers}

\begin{figure}
  \centering
  \includegraphics[width=0.7\textwidth]{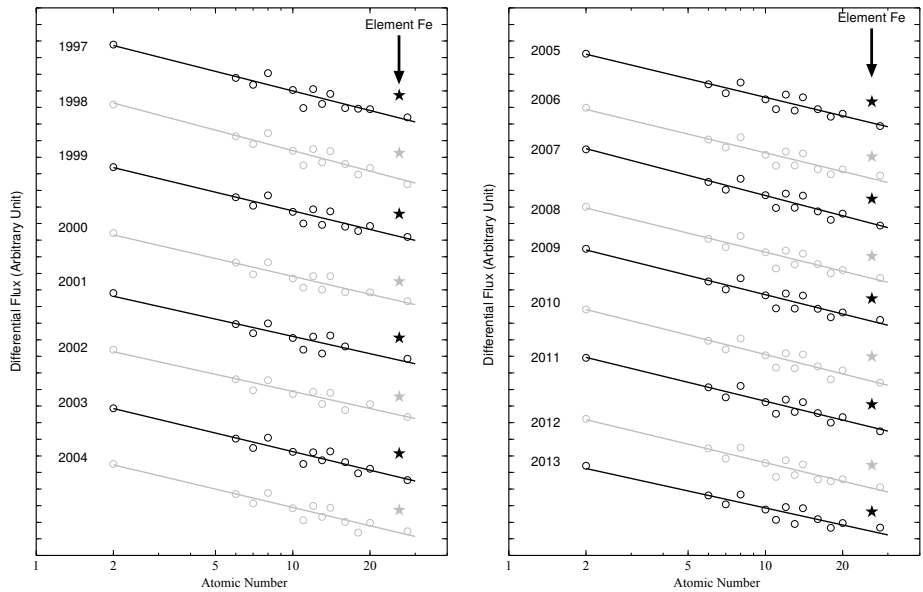}
  \caption{Fitted $30$ MeV/nuc element fluxes for yearly average of GCRs are shown
as a function of atomic numbers. The results are multipled by a free number for 
presentation purpose. The solid lines are the least-square linear fit in log-log 
space. The solid star indicates the element Fe and are not included in the linear
fit.}
  \label{fig:gcr}
  \end{figure}

Figure \ref{fig:gcr} shows the fitted $30$ MeV/nuc GCR flux as a function of element
 numbers for each year from 1997 to 2013. Note that the results are multiplied by
a free parameter for the purpose of presentation. From the figure we can see that 
the GCR fluxes can be linearly fitted in log-log space with slopes, intercepts, and 
correlation coefficients (Corr. Coef.). Note that in Figure \ref{fig:gcr} the filled
 stars indicate Fe which is not included in the power law fitting (solid line) to
remove the iron peak.

\begin{figure}
  \centering
  \includegraphics[width=0.7\textwidth]{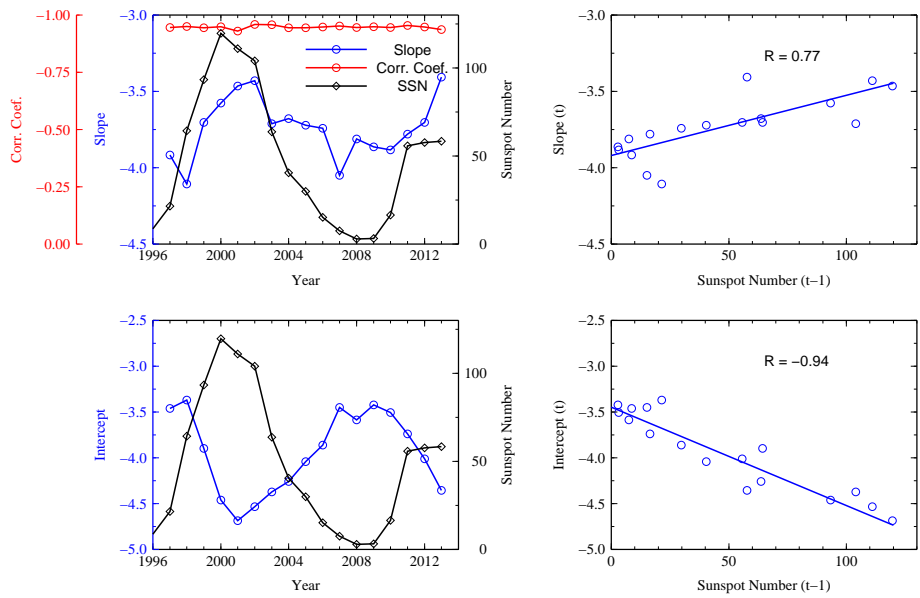}
  \caption{Similar as in Figure \ref{fig:slope}, the fitting results in Figure
\ref{fig:gcr} are shown, except in top right panel the Corr. Coef. are not shown.}
  \label{fig:slope-gcr}
  \end{figure}

Similar as in Figure \ref{fig:slope}, the fitted slopes, intercepts, and correlation
 coefficients in Figure 
\ref{fig:gcr} is shown in Figure \ref{fig:slope-gcr}. We can see the Corr. Coef. of 
the fitting results of GCRs in Figure \ref{fig:gcr} is nearer to $-1$ than that
of SEPs in Figure \ref{fig:fip}, indicating better linear fitting of GCRs. Note that
in order to linear-fit the slopes and intercepts from the fitting results in Figure
\ref{fig:gcr}, the SSN is one year delayed because there is such a delay of solar
modulation for GCRs in heliosphere.

\section{Shannon Entropy of SEPs and CRs}

Following \cite{Laurenza2012}, we 
use Shannan enertropy (information entropy) to investigate the spectrum evolution of
 SEPs and CRs.
Firstly, the entropy $S$ can be written as
\begin{linenomath}
\begin{equation}
  S=-k\sum_{i=1}^{N}p_i\ln p_iE_i,
  \label{equ:entropy}
\end{equation}
\end{linenomath}
where $E_i$ denotes the $i$th energy interval, $f_i$ denotes the corresponding 
flux, $p_i=f_i/\sum\limits_{i=1}^{N}f_i E_i$ denotes the corresponding probability
density, and $k$ is a constant which is generally assumed as $k=1/\log(N)$.

\begin{figure}
  \centering
  \includegraphics[width=.7\textwidth]{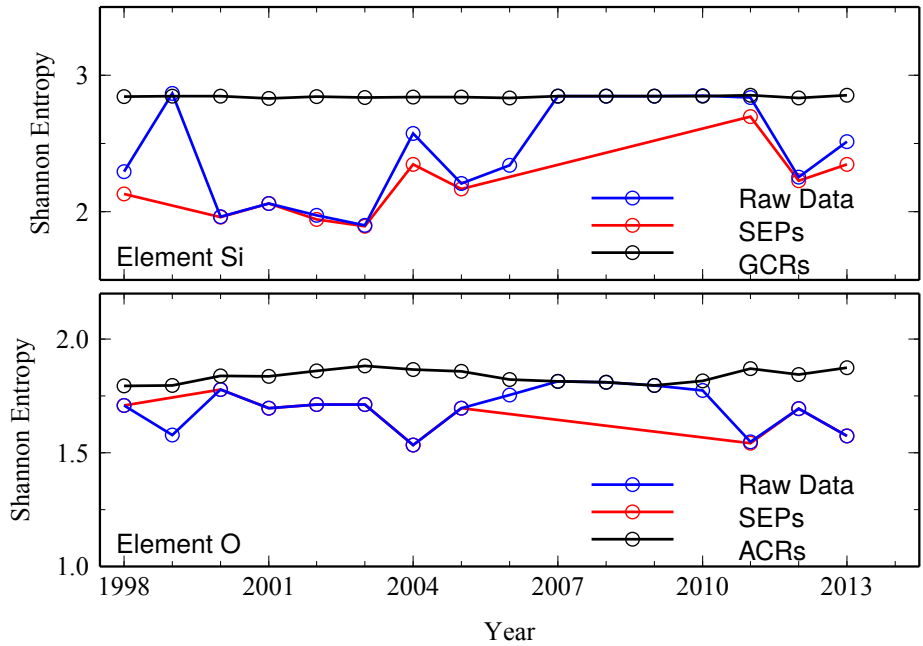}
  \caption{The entropy of element Si with all ACE/SIS energy channels ($11.0$
MeV/nuc - $103.6$ MeV/nuc) and that of element O with the first four ACE/SIS energy 
channels
($<20$ MeV/nuc) over the period from the year 1998 to 2013. The entropy of raw data,
 SEPs and CRs are denoted as blue, red and black lines, respectively. The entropy of
 SEPs in the years 1999, and 2006-2010 are not included.}
  \label{fig:entropy}
\end{figure}

Next, we calculate the yearly averaged ACE/SIS flux of element Si with all energy
channels ($11.0$ MeV/nuc - $103.6$ MeV/nuc) and that of O with the first four energy
 channels ($<20$ MeV/nuc), for raw data, CRs (background data), and SEPs (spike 
data). 
Top panel of Figure \ref{fig:entropy} shows entropy of the element Si, and bottom 
panel shows entropy of the element O. In the figure, blue, red, and black lines 
indicate the raw data, the SEPs (spike data), and the CRs (background data),
respectively. Similar as above, we do not include the SEPs data in the years 1999,
and 2006-2010.
It is noted that in top panel of Figure \ref{fig:entropy} the CRs data
 of Si are GCRs, and that in the bottom panel the CRs data of O are mostly ACRs.
From the figure we can see, the entropy of GCRs of Si is almost a constant but that 
of the ACRs of O has weak solar cycle variations. In addition, the entropy of Si and
 O SEPs are smaller than that of Si GCRs and O ACRs, respectively. Furthermore, the 
entropy of raw data usually closely track that of SEPs, but during the recent 
extreme solar minimum, the entropy of raw data is almost identical to that of CRs 
because the raw data are dominated with CRs.

\section{Discussion and Conclusion}

In this work, we use a despiking method to automatically separate CRs, including
GCRs and ACRs, from SEPs over the period more than one solar cycle for ACE/SIS heavy
 ion measurements, so that we are able to study solar cycle variations of CRs and
SEPs in the same energy ranges simultaneously with a computer code. Finally, we have
 the following findings.

It is shown that CRs usually show solar cycle variations with different levels of
strength according to different energy and elements. In addition, although SEPs show
 much larger variations in any epoch of solar activity, they still show moderate 
correlation  with solar cycles. The possible reason is that during solar maximum and
 descending phases, there are more large solar events and in solar events there are
 more enhancements of seed particles because of more preceding solar events, so the 
average SEP fluxes are larger than that during solar minimum and 
ascending phases. Therefore, different elemental ratios, Fe/O or N/O, for SEPs or
CRs in different energy range show different solar cycle variations.

In addition, it is shown that the SEP fluxes relative to the recent photospheric 
abundances of elements can be linearly fit with FIP in log-log space to get fitting
slopes, correlation coefficients, and intercepts. It is also shown that both slopes 
and correlation coefficients show strong anti-correlation with SSN and that
intercepts show correlation with SSN when there are enough SEP events for 
statistics. However, for the same elements CRs behave much differently. In each 
year, CR fluxes can be fit linearly with atomic numbers in log-log space. In 
addition, the fitting slopes and intercepts can also be fitted linearly with SSN
with one year delay.

Furthermore, we find that for the same element in the same energy range, the 
variations of Shannon entropy of GCRs are much smaller than that of the ACRs 
entropy, and the SEPs entropy is usually smaller than that of CRs. It is known that 
when the uncertainty of a system is increased, the Shannon entropy is increased. And
 also since the SEPs are associated with sloar eruptions with evolution in days, but
 CRs have much longer time of evolution, so CRs have much larger uncertainty than 
SEPs. Therefore, SEPs usually have smaller value of shannon entropy than CRs. In 
addition, the sources of ACRs are inside heliosphere and modified by solar activity,
but the sources of GCRs are outside heliosphere,
so the solar cycle variation of ACRs entropy is larger than GCRs entropy.

The above results from data analysis are useful for us to understand different 
compositions, origins, and modulation of CRs (GCRs and ACRs) and SEPs. However, in 
order to deepen our understanding of CRs and SEPs' characteristics, we would combine
 the data analysis and the CR 
modulation models \citep[e.g.,][]{ZhaoEA2013} and SEP transport models 
\citep[e.g.,][]{Qin2006JGR} to numerically study the variations of 
CRs and SEPs over solar cycles. 
%

\begin{acknowledgments}
We are partly supported by grants NNSFC 41374177, NNSFC 41125016.
The data sets of SIS and CRIS are downloaded from the ACE Science Center
archives at http://www.srl.caltech.edu/ACE/ASC/level2/. We also acknowledge
the use of the sunspot number data provided to the community by NASA.
\end{acknowledgments}



\end{article}
\clearpage
\begin{table}[p]
\caption{Recommended energy in unit of MeV/nuc of each measurement energy intervals 
of ACE/SIS obtained from ACE Home page (http://www.srl.caltech.edu/ACE/ASC/). 
\label{tbl:energy_interval}}
\begin{flushleft}
\begin{tabular*}{\textwidth}{@{\extracolsep{\fill}}|c|c|c|c|c|c|c|c|c|}
\tableline
Element  & $E_1$ & $E_2$ & $E_3$ & $E_4$ & $E_5$ & $E_6$ & $E_7$ & $E_8$\\
\hline
$He(2)$ & $4.0$ & $5.4$ & $6.7$ & $8.4$ & $11.5$ & $15.6$ & $23.0$ & $34.8$\\
$C(6)$ & $7.4$ & $9.8$ & $12.3$ & $15.5$ & $21.2$ & $28.9$ & $42.5$ & $64.4$ \\
$N(7)$ & $8.0$ & $10.7$ & $13.3$ & $16.9$ & $23.1$ & $31.4$ & $46.3$ & $70.2$ \\
$O(8)$ & $8.5$ & $11.4$ & $14.3$ & $18.1$ & $24.8$ & $33.8$ & $49.8$ & $75.7$ \\
$Ne(10)$ & $9.5$ & $12.8$ & $16.0$ & $20.4$ & $28.0$ & $38.2$ & $56.4$ & $85.8$ \\
$Na(11)$ & $9.6$ & $13.0$ & $16.4$ & $20.8$ & $28.7$ & $39.3$ & $58.0$ & $88.2$\\
$Mg(12)$ & $10.3$ & $14.0$ & $17.6$ & $22.4$ & $30.9$ & $42.2$ & $62.4$ & $95.0$\\
$Al(13)$ & $10.4$ & $14.1$ & $17.8$ & $22.8$ & $31.5$ & $43.1$ & $63.8$ & $97.3$\\
$Si(14)$ & $11.0$ & $15.0$ & $18.9$ & $24.2$ & $33.4$ & $45.8$ & $67.9$ & $103.6$\\
$S(16)$ & $11.6$ & $15.9$ & $20.2$ & $25.8$ & $35.8$ & $49.2$ & $73.0$ & $111.7$\\
$Ar(18)$ & $12.1$ & $16.9$ & $21.5$ & $27.7$ & $38.5$ & $53.2$ & $79.1$ & $121.2$\\
$Ca(20)$ & $12.6$ & $17.5$ & $22.4$ & $28.8$ & $40.1$ & $55.3$ & $82.4$ & $126.6$\\
$Fe(26)$ & $13.0$ & $18.5$ & $23.8$ & $30.9$ & $43.5$ & $60.6$ & $90.9$ & $140.4$\\
$Ni(28)$ & $13.7$ & $19.5$ & $25.2$ & $32.7$ & $46.2$ & $64.4$ & $96.7$ & $149.7$\\
\tableline
\end{tabular*}
\end{flushleft}
\end{table}

\clearpage
\clearpage
\begin{table}[p]
\caption{Energy channels shown in Table \ref{tbl:energy_interval} for elements N, O,
 and Fe. \label{tbl:channels}}
\begin{flushleft}
\begin{tabular*}{.5\textwidth}{@{\extracolsep{\fill}}ccccc}
\tableline
\multirow{2}{*}{Channel} & \multirow{2}{*}{$E_t$(MeV/nuc)} 
&\multicolumn{3}{c}{$E_i$}
\\
\cline{3-5}
        &              &Fe&Ni&O\\ 
\tableline
I  & $13$ & $E_1$ & $E_1$ & $E_3$\\
\tableline
II & $31$ & $E_4$ & $E_4$ & $E_6$\\
\tableline
III& $46$ & $E_5$ & $E_5$ & $E_7$\\
\tableline
\end{tabular*}
\end{flushleft}
\end{table}
\clearpage

\clearpage

\clearpage

\clearpage

\clearpage

\clearpage

\clearpage

\clearpage

\clearpage

\clearpage

\clearpage

\end{document}